\documentclass[journal]{IEEEtran}
\usepackage{graphicx}
\usepackage{color}
\usepackage{placeins}
\usepackage{float}
\usepackage{tabularx,colortbl}
\usepackage{graphics, theorem, times, amsfonts, amsmath, amssymb, cite, verbatim}
\usepackage{multicol,multirow,booktabs}
\usepackage{tikz}
\usepackage{framed}
\usetikzlibrary{shapes,arrows}
\usepackage{pgfplots}
\usepackage{color}
\usepackage{hyperref}

\usepackage{needspace}

\newcounter{excercise}
\newcounter{excercisepart}

\definecolor{pennblue}{cmyk}{1,0.65,0,0.30}
\definecolor{pennred}{cmyk}{0,1,0.65,0.34}
\definecolor{mygreen}{rgb}{0.10,0.50,0.10}

\def \diag    {\text{\normalfont diag} }

\def \reals    {{\mathbb R}}

\def\mbI{{\ensuremath{\mathbb I}}}

\def\ccalG{{\ensuremath{\mathcal G}}}

\def\ccalU{{\ensuremath{\mathcal U}}}
\def\ccalV{{\ensuremath{\mathcal V}}}

\def\ccal0{{\ensuremath{\mathcal 0}}}
\def\tdh{{\ensuremath{\tilde h}}}

\def\tdx{{\ensuremath{\tilde x}}}

\def\bbD{{\ensuremath{\mathbf D}}}

\def\bbH{{\ensuremath{\mathbf H}}}
\def\bbI{{\ensuremath{\mathbf I}}}

\def\bbL{{\ensuremath{\mathbf L}}}
\def\bbM{{\ensuremath{\mathbf M}}}

\def\bbW{{\ensuremath{\mathbf W}}}

\def\bbV{{\ensuremath{\mathbf V}}}

\def\bbh{{\ensuremath{\mathbf h}}}

\def\bbv{{\ensuremath{\mathbf v}}}

\def\bbt{{\ensuremath{\mathbf t}}}
\def\bbx{{\ensuremath{\mathbf x}}}

\def\bb0{{\ensuremath{\mathbf 0}}}
\def\tbH{{\tilde{\ensuremath{\mathbf H}} }}

\def\tbh{{\tilde{\ensuremath{\mathbf h}} }}

\def\tbx{{\tilde{\ensuremath{\mathbf x}} }}

\def\bbLambda{\boldsymbol{\Lambda}}

\def\bbPsi{\boldsymbol{\Psi}}

\def \lam    {\lambda}

\def \bbLam  {\bbLambda}
\newtheorem{remark}{\hspace{0pt}\bf Remark}

\newtheorem{definition}{\hspace{0pt}\bf Definition}

\def \L {\text{L}}
\def \M {\text{M}}
\def \HH{\text{H}}

\def \TV{\text{TV}}
\def \ZC{\text{ZC}}

\def \visual{\text{visual}}
\def \motor{\text{motor}}
\def \other{\text{others}}
\def \vm{\text{visual-motor}}
\def \vo{\text{visual-others}}
\def \mo{\text{motor-others}}

\begin{document}

\title{Graph Frequency Analysis of Brain Signals}

%
\author{Weiyu Huang, Leah Goldsberry, Nicholas F. Wymbs, Scott T. Grafton, Danielle S. Bassett and Alejandro Ribeiro
\thanks{Supported by NSF CCF-1217963, PHS NS44393, ARO ICB W911NF-09-0001, ARL W911NF-10-2-0022, ARO W911NF-14-1-0679, NIH R01-HD086888, NSF BCS-1441502, NSF BCS-1430087 and the  John D. and Catherine T. MacArthur and Alfred P. Sloan foundations. Content does not necessarily represent official views of any of the funding agencies. W. Huang, L. Goldsberry, D. S. Bassett, and A. Ribeiro are with the Dept. of Electrical and Systems Eng., University of Pennsylvania. N. F. Wymbs is with the Dept. of Physical Medicine and Rehabilitation, Johns Hopkins University. S. T. Grafton is with the Dept. of Psychological and Brain Sciences, University of California at Santa Barbara. Email: whuang@seas.upenn.edu, goleah@seas.upenn.edu, nwymbs1@jhmi.edu, scott.grafton@psych.ucsb.edu, dsb@seas.upenn.edu, and aribeiro@seas.upenn.edu.}}

\maketitle

\begin{abstract} 
This paper presents methods to analyze functional brain networks and signals from graph spectral perspectives. The notion of frequency and filters traditionally defined for signals supported on regular domains such as discrete time and image grids has been recently generalized to irregular graph domains, and defines brain graph frequencies associated with different levels of spatial smoothness across the brain regions. Brain network frequency also enables the decomposition of brain signals into pieces corresponding to smooth or rapid variations. We relate graph frequency with principal component analysis when the networks of interest denote functional connectivity. The methods are utilized to analyze brain networks and signals as subjects master a simple motor skill. We observe that brain signals corresponding to different graph frequencies exhibit different levels of adaptability throughout learning. Further, we notice a strong association between graph spectral properties of brain networks and the level of exposure to tasks performed, and recognize the most contributing and important frequency signatures at different task familiarity.
\end{abstract}

\IEEEpeerreviewmaketitle

%
\section{Introduction}\label{sec_introduction}

The study of brain activity patterns has proven valuable in identifying neurological disease and individual behavioral traits \cite{haken2013, fox2007, britz2010, cole2006}. The use of functional brain networks describing the tendency of different regions to act in unison has proven complementary in the analysis of similar matters \cite{medaglia2016, achard2006, bullmore2012, richiardi2013}. It is not surprising that signals and networks prove useful in similar problems since the two are closely related. In this paper we advocate an intermediate path in which we interpret brain activity as a signal supported on the graph of brain connectivity. We show how the use of graph signal processing (GSP) tools can be used to glean information from brain signals using the network as an aid to identify patterns of interest. The benefits of incorporating network information into signal analysis has been demonstrated in multiple domains. Notable examples of applications include video compression \cite{nguyen2014}, rating predictions in recommendation systems \cite{ma2015}, and breast cancer diagnostics \cite{SegarraEtal15}, and semi-supervised learning \cite{gadde2014}.

The fundamental GSP concepts that we utilize to exploit brain connectivity in the analysis of brain signals are the graph Fourier transform (GFT) and the corresponding notions of graph frequency components and graph filters. These concepts are generalizations of the Fourier transform, frequency components, and filters that are used in regular domains such as time and spatial grids \cite{sandryhaila2013, sandryhaila2014, shuman2013}. As such, they permit the decomposition of a graph signal into components that represent different modes of variability. We can define low graph frequency components representing signals that change slowly with respect to brain connectivity networks in a well defined sense and high graph frequency components representing signals that change fast in the same sense. This is important because low and high {\it temporal} variability have proven important in the analysis of neurological disease and behavior \cite{garrett2012, Heisz2012}. GFT based decompositions permit a similar analysis of variability across regions of the brain for a fixed time -- a sort of {\it spatial} variability measured with respect to the connectivity pattern. We demonstrate here that it is useful in a similar sense; see e.g. Figs. \ref{fig_Signal_6Week}, \ref{fig_Signal_3Day}, \ref{fig_GFT_Mag_Sub}, and \ref{fig_Freq_LR}. 

The GSP studies in this paper are related to principal component analysis (PCA), which has been used with success in the analysis of brain signals \cite{leonardi2013, viviani2005}. The difference with the GSP analysis we present here is that PCA implicitly assumes the brain network to be a correlation matrix and the signals to be drawn from a stochastic model. More importantly, whereas the GFT can be used to, e.g., decompose the signal into low, medium, and high frequency components, PCA is mostly utilized for dimensionality reduction; which in the language of this paper is tantamount to analyzing a few low graph frequency components. Another important difference is that PCA focuses on identifying variability across {\it different} realizations of brain signals, but the GFT identifies spatial variability of a {\it single} realization. GSP is also related to the spectral analysis of networks in general and Laplacians in particular \cite{harrison2008, shi2000}. The difference in this case is that these spectral analyses yield properties of the networks. In GSP analyses, the network provides an underlying structure, but the interest is on signals expressed on this stratum. 

The goal of this paper is to introduce GSP notions that can be used to analyze brain signals and to demonstrate their value in identifying patterns that appear when monitoring activity as subjects learn to perform a visual-motor task. Specifically, the contributions of this paper are: (i) To explain tools from the emerging field of GSP and show how they can be applied in analyzing brain signals. (ii) To evaluate the graph spectrum of brain functional network and to define artificial network construction methods that replicate the features of the graph spectrum of functional networks. (iii) To examine the temporal variation of brain signals corresponding to different graph frequencies when participants perform visual-motor learning tasks. (iv) To investigate the contribution of brain signals associated with different graph frequencies to the learning success at different stages of visual-motor learning. 

We begin the paper with the introduction of basic notions of graphs and graph signals. Particular emphasis goes into the definition of the graph Fourier transform and the interpretation of graph frequency components as different modes of spatial variability measured with respect to the brain network (Section \ref{sec_graph_frequency}). We also introduce the notion of graph filters and discuss the interpretation of a graph low-pass filter as a local averaging operation (Section \ref{sec_decomposition}). Band-pass and high-pass filters that are used in later sections to separate signals into different components are also introduced. We point out that the discussion here is more extensive than necessary for readers familiar with GSP. The goal is to make the paper accessible to readers that are not necessarily familiar with the subject. 

We then move on to describe two different experiments involving the learning of different visual-motor tasks by different sets of participants (Section \ref{sec_dataset}). The graph frequency of functional brain networks for the participants is visualized and analyzed (Section \ref{sec_network_illustration}), which is found to be associated across scan sessions in the same dataset and across datasets. In specific, we find that graph high frequencies of functional networks concentrate on visual and sensorimotor modules of the brain -- the two brain areas well-known to be associated with motor learning \cite{kleim2002, bassett2015}. This motivates us to consider graph frequencies besides low frequency components, whereas the PCA-oriented approaches has been focusing on low frequencies. We also describe the construction of a simple model to establish artificial networks with a few network descriptive parameters (Section \ref{sec_artificial_network}). We observe that the model is able to mimic the properties of actual functional brain networks and we use them to analyze spectral properties of the brain networks (Section \ref{sec_spectral_property}). The paper then utilizes graph frequency decomposition to visualize and investigate brain activities with different levels of spatial variation (Section \ref{sec_signal_decomp}). It is noticed that the decomposed signals associated to different graph frequencies exhibit different levels of temporal variation throughout learning (Section \ref{sec_variance}). In particular, graph low and high frequency components exhibit higher temporal variation at multiple temporal scales in the two experiments considered. Because temporal variation has been shown to be associated with learning success \cite{Heisz2012}, this implies graph low and high frequency components offer more contributions to learning success. Finally, we also define learning capabilities of subjects, and examine the importance of brain frequencies at different task familiarity by evaluating their respective correlation with learning performance at different task familiarities (Section \ref{sec_learning}). We find that it is good to have graph low frequency components (smooth, spread, and cooperative brain signals) when we face an unfamiliar task. When we become highly familiar with the task, it is better and favors further learning to have graph high frequency components (varied, spiking, and competitive brain signals).

%
\section{Graph Signal Processing}\label{sec_gsp}

The interest of this paper is to study brain signals in which we are given a collection of measurements $x_i$ associated with each cortical region out of $n$ different brain regions. An example signal of this type is an fMRI reading in which $x_i$ estimates the level of activity of brain region $i$. The collection of $n$ measurements is henceforth grouped in the vector signal $\bbx = [x_1, x_2, \ldots, x_n]^T \in \reals^n$. A fundamental feature of the signal $\bbx$ is the existence of an underlying pattern of structural or functional connectivity that couples the values of the signal $\bbx$ at different brain regions. Irrespective of whether connectivity is functional or structural, our goal here is to describe tools that utilize this underlying brain network to analyze patterns in the neurophysiological signal $\bbx$.

We do so by modeling connectivity between brain regions with a network that is connected, weighted, and symmetric. Formally, we define a network as the pair $\ccalG=(\ccalV, \bbW)$, where $\ccalV=\{1,\ldots,n\}$ is a set of $n$ vertices or nodes representing individual brain regions and $\bbW \in \reals^{n \times n}$ collects weights of edges in the network with $w_{ij}\geq0$ the weight of the edge $(i,j)$. Since the network is undirected and symmetric we have that $w_{ij}=w_{ji}$ for all $(i,j)$. The weights $w_{ij}=w_{ji}$ represent the strength of the connection between regions $i$ and $j$, or, equivalently, the proximity or similarity between nodes $i$ and $j$. In terms of the signal $\bbx$, this means that when the weight $w_{ij}$ is large, the signal values $x_i$ and $x_j$ tend to be related. Conversely, when the weight $w_{ij}$ is small, 
the signal values $x_i$ and $x_j$ are not directly related except for what is implied by their separate connections to other nodes. 

We adopt the conventional definitions of the degree and Laplacian matrices \cite[Chapter 1]{Chung97}. The degree matrix $\bbD \in \reals_+^{n\times n}$ is a diagonal matrix with its $i$th diagonal element $D_{ii}=\sum_j w_{ij}$ denoting the sum of all the weights of edges out of node $i$. The Laplacian matrix is defined as the difference $\bbL:=\bbD-\bbW\in\reals^{n\times n}$. The components of the Laplacian matrix are explicitly given by $ L_{ij} = -w_{ij}$ and $L_{ii} = \sum_{j=1}^n w_{ij}$. Observe that the Laplacian is real, symmetric, diagonal dominant, and with strictly positive diagonal elements. As such, 
the matrix $\bbL$ is positive semidefinite. 
 The eigenvector decomposition of $\bbL$ is utilized in the following section to define the graph Fourier transform and the associated notion of graph frequencies.

We note that brain networks, irrespective of whether their connectivity is functional \cite{zhang2003} or structural \cite{grutzendler2002}, tend to be stable for a window of time, entailing associations between brain regions during captured time of interest. Brain activities can vary more frequently, forming multiple samples of brain signals supported on a common underlying network.

%
\subsection{Graph Fourier Transform and Graph Frequencies}\label{sec_graph_frequency}

Begin by considering the set $\{\lambda_k\}_{k = 0, 1, \dots, n - 1}$ of eigenvalues of the Laplacian $\bbL$ and assume they are ordered so that $0=\lam_0\leq\lam_1\leq\ldots\leq\lam_{n-1}$. Define the diagonal eigenvalue matrix $\bbLam:=\diag(\lam_0,\lam_1\ldots,\lam_{n-1})$ and the eigenvector matrix 
\begin{align}\label{eqn_eigenvector_matrix}
   \bbV := [\bbv_0,\bbv_1,\ldots,\bbv_{n-1}].
\end{align}
Because the graph Laplacian $\bbL$ is real symmetric, it accepts the eigenvalue decomposition
\begin{align}\label{eqn_eigenvalue_decomposition}
   \bbL =  \bbV \bbLambda \bbV^{H},
\end{align}
where $\bbV^{H}$ represents the Hermitian (conjugate transpose) of the matrix $\bbV$. The validity of \eqref{eqn_eigenvalue_decomposition} follows because the eigenvectors of symmetric matrices are orthogonal so that the definition in \eqref{eqn_eigenvector_matrix} implies that $\bbV^{H} \bbV = \bbI$. The eigenvector matrix $\bbV$ is used to define the Graph Fourier Transform of the graph signal $\bbx$ as we formally state next; see, e.g., \cite{shuman2013}.

\begin{definition}\label{def_gft} Given a signal $\bbx\in\reals^n$ and a graph Laplacian $\bbL\in\reals^{n\times n}$ accepting the decomposition in \eqref{eqn_eigenvalue_decomposition}, the Graph Fourier Transform (GFT) of $\bbx$ with respect to $\bbL$ is the signal 
\begin{align}\label{eqn_GFT}
   \tbx = \bbV^H \bbx .
\end{align}
The inverse (i)GFT of $\tbx$ with respect to $\bbL$ is defined as 
\begin{align}\label{eqn_iGFT}
    \bbx = \bbV \tbx.
\end{align}
We say that $\bbx$ and $\tbx$ form a GFT transform pair. 
\end{definition}

Observe that since $\bbV\bbV^H=\bbI$, the iGFT is, indeed, the inverse of the GFT. Given a signal $\bbx$ we can compute the GFT as per \eqref{eqn_GFT}. Given the transform $\tbx$ we can recover the original signal $\bbx$ through the iGFT transform in 
\eqref{eqn_iGFT}.

There are several reasons that justify the association of the GFT with the Fourier transform. Mathematically, it is just a matter of definition that if the vectors $\bbv_k$ in \eqref{eqn_eigenvalue_decomposition} are of the form $\bbv_k = [1, e^{j2\pi k/n}, \ldots, e^{j2\pi k(n-1)/n}]^T$, the GFT and iGFT in Definition \ref{def_gft} reduce to the conventional time domain Fourier and inverse Fourier transforms. More deeply, it is not difficult to see that if the graph $\ccalG$ is a cycle, the vectors $\bbv_k$ in \eqref{eqn_eigenvalue_decomposition} are of the form $\bbv_k = [1, e^{j2\pi k/n}, \ldots, e^{j2\pi k(n-1)/n}]^T$. Since cycle graphs are representations of discrete periodic signals, it follows that the GFT of a time signal is equivalent to the conventional discrete Fourier transform; see, e.g., \cite{Segarra2015b}.

An important property of the GFT is that it encodes a notion of variability akin to the notion of variability that the Fourier transform encodes for temporal signals. To see this, define $\tbx=[\tdx_0,\ldots,\tdx_{n-1}]^T$ and expand the matrix product in \eqref{eqn_iGFT} to express the original signal $\bbx$ as
\begin{align}\label{eqn_iGFT_expanded}
    \bbx = \sum_{k=0}^{n-1} \tdx_k \bbv_k.
\end{align}
It follows from \eqref{eqn_iGFT_expanded} that the iGFT allows us to write the signal $\bbx$ as a sum of orthogonal components $\bbv_k$ in which the contribution of $\bbv_k$ to the signal $\bbx$ is the GFT component $\tdx_k$. In conventional Fourier analysis, the eigenvectors $\bbv_k = [1, e^{j2\pi k/n}, \ldots, e^{j2\pi k(n-1)/n}]^T$ carry a specific notion of variability encoded in the notion of frequency. When $k$ is close to zero, the corresponding complex exponential eigenvectors are smooth. When $k$ is close to $n$, the eigenfunctions fluctuate more rapidly in the discrete temporal domain. In the graph setting, the graph Laplacian eigenvectors provide a similar notion of frequency. Indeed, define the total variability of the graph signal $\bbx$ with respect to the Laplacian $\bbL$ as
\begin{align}\label{eqn_TV_general}
    \text{TV}(\bbx) = \bbx ^H \bbL \bbx  = \sum_{i \neq j} w_{ij} (x_i-x_j)^2,
\end{align}
where in the second equality we expanded the quadratic form. It follows that the total variation $\text{TV}(\bbx)$ is a measure of how much the signal changes with respect to the network. For the edge $(i,j)$, when $w_{ij}$ is large we expect the values $x_i$ and $x_j$ to be similar because a large weight $w_{ij}$ is encoding functional similarity between brain regions $i$ and $j$. The contribution of their difference $(x_i - x_j)^2$ to the total variation is amplified by the weight $w_{ij}$. If the weight $w_{ij}$ is small, activities at brain regions $i$ and $j$ tend to be uncorrelated, and therefore the difference between the signal values $x_i$ and $x_j$ makes little contribution to the total variation. 
We can then think of a signal with small total variation as one that changes slowly over the graph and of signals with large total variation as those that change rapidly over the graph.

Consider the total variation of the eigenvectors $\bbv_k$ and use the facts that $\bbL \bbv_k = \lambda_k\bbv_k$ and $\bbv_k^H\bbv_k=1$ to conclude that 
\begin{align}\label{eqn_TV}
    \text{TV}(\bbv_k) = \bbv_k^H \bbL \bbv_k = \lam_k.
\end{align}
It follows from \eqref{eqn_TV} and the fact that the eigenvalues are ordered as $0=\lam_0\leq\ldots\leq\lam_{n-1}$, that the total variations of the eigenvectors $\bbv_k$ follow the same order. Combining this observation with the discussion following \eqref{eqn_TV_general}, we conclude that when $k$ is close to $0$, the eigenvectors $\bbv_k$ vary slowly over the graph, whereas for $k$ close to $n$ the eigenvalues vary more rapidly. Therefore, from \eqref{eqn_iGFT_expanded} we see that the GFT and iGFT allow us to decompose the brain signal $\bbx$ into components that characterize different levels of variability. The GFT coefficients $\tdx_k$ for small values of $k$ indicate how much these slowly varying signals contribute to the observed brain signal $\bbx$. On the other hand, the GFT coefficients $\tdx_k$ for large values of $k$ describe how much rapidly varying signals contribute to the observed brain signal $\bbx$.

%
\subsection{Graph Filtering and Frequency Decompositions}\label{sec_decomposition}

Given a graph signal $\bbx$ with GFT $\tbx$ we can isolate the frequency components corresponding to the lowest $K_\L$ graph frequencies by defining the filtered spectrum $\tbx_\L = \tbH_\L \tbx$ satisfying $\tdx_{\L k} = \tdx_k$ for $k<K_L$ and $\tdx_{\L k}=0$ otherwise. The filter $\tbH_\L$ can be written as the diagonal matrix $\tbH_\L = \diag(\tbh_\L)$ where the vector $\tbh_\L$ takes value 1 for frequencies smaller than $K_\L$ and is otherwise null,
\begin{align}\label{eqn_low_pass_graph_filter}
    \tdh_{\L k} = \mbI\big[k < K_\L \big] .
\end{align}
Utilizing the definitions of the GFT in \eqref{eqn_GFT} and the iGFT in \eqref{eqn_iGFT}, the spectral operation $\tbx_\L = \tbH_\L \tbx$ is equivalent to performing the following operations in the graph vertex domain
\begin{align}\label{eqn_low_pass_frequency}
     \bbx_\L\ :=\ \bbV \tbx_\L 
            \  =\ \bbV \tbH_\L \tbx 
            \  =\ \bbV \tbH_\L \bbV^{-1} \bbx 
            \ :=\ \bbH_\L \bbx.
\end{align}
The last equality in \eqref{eqn_low_pass_frequency} defines the matrix $\bbH_\L := \bbV \tbH_\L \bbV^{-1}$ so that the graph signal $\bbx_\L$ associated with low graph frequencies of $\bbx$ can be written as the product $\bbx_\L = \bbH_\L \bbx$. Since the signal $\bbx_\L$ contains the low graph frequency components of $\bbx$, we say the matrix $\bbH_\L$ in \eqref{eqn_low_pass_frequency} is a {\it graph low-pass filter}. 

The filter $\bbH_\L := \bbV \tbH_\L \bbV^{-1}$ admits an alternative representation as the expansion $\bbH_\L = \sum_{k = 0}^{n-1} h_{\L k} \bbL^k$ in terms of Laplacian powers \cite{Segarra2015b}. The coefficients $h_{\L k}$ in this expansion are elements of the vector $\bbh_\L = \bbPsi^{-1} \tbh_\L$ where $\bbPsi$ is the Vandermonde matrix defined by the eigenvalues of $\bbL$, i.e.,
\begin{align}\label{eqn_Vandermonde}
    \bbPsi = \begin{bmatrix} 
                   1      & \lambda_0     & \cdots & \lambda_0^{n-1}     \\
                   \vdots & \vdots        & \ddots & \vdots              \\
                   1      & \lambda_{n-1} & \cdots & \lambda_{n-1}^{n-1} 
             \end{bmatrix} .
\end{align}
Since the eigenvalues are ordered in \eqref{eqn_Vandermonde}, the coefficients $h_{\L k}$ tend to be concentrated in small indexes $k$, and the expansion $\bbH_\L = \sum_{k = 0}^{n-1} h_{\L k} \bbL^k$ is therefore dominated by small powers $\bbL^k$. From this fact it follows that we can think of the graph low-pass filtered signal $\bbx_\L$ as resulting from a localized averaging of the elements of $\bbx$. To understand this interpretation, simply note that $\bbL^0\bbx = \bbx$ coincides with the original signal, $\bbL\bbx$ is an average of neighboring elements, $\bbL^2\bbx$ is an average of elements in nodes that interact via intermediate common neighbors, and, in general, $\bbL^k\bbx$ describes interactions between $k$-hop neighbors. The fact that $\bbx_\L$ can be considered as a signal that follows from local averaging of $\bbx$ implies that $\bbx_\L$ has smaller total variation than $\bbx$ and is consistent with the interpretation of low graph frequencies presented in Section \ref{sec_graph_frequency}. We point out that the definition $\bbh_\L = \bbPsi^{-1} \tbh_\L$ assumes the inverse matrix $\bbPsi^{-1}$ exists. This holds true if the graph Laplacian does not have duplicate eigenvalues, which is the case for all functional brain networks examined in the paper.

Other types of graph filters can be defined analogously to study interactions between signal components other than the local interactions captured in $\bbx_\L$. Apart from the graph low-pass filter $\bbH_\L$, we also consider a graph band-pass filter $\bbH_\M$ and a graph high-pass filter $\bbH_\HH$, whose graph frequency responses are defined as
\begin{alignat}{2}
    &\tdh_{\M k}  &&= \mbI\big[K_\L \leq k < K_\L + K_\M \big],\label{eqn_filter_decomp_M}\\
    &\tdh_{\HH k} &&= \mbI\big[K_\L + K_\M \leq k \big]       .\label{eqn_filter_decomp_H}
\end{alignat}
The definitions in \eqref{eqn_low_pass_graph_filter}, \eqref{eqn_filter_decomp_M}, and \eqref{eqn_filter_decomp_H} are such that the low-pass filter takes the lowest $K_\L$ graph frequencies, the band-pass filter captures the middle $K_\M$ graph frequencies, and the high-pass filter the highest $n - K_\L - K_\M$ frequencies. The three filters are defined such that the graph frequencies of their respective interest are mutually exclusive yet collectively exhaustive. As a result, if we use $\bbx_\M := \bbH_\M\bbx$ and $\bbx_\HH:=\bbH_\HH\bbx$ to respectively denote the signals filtered by the band-pass and high-pass filters, we have that the original signal can be written as the sum $\bbx = \bbx_\L + \bbx_\M + \bbx_\HH$. This gives a decomposition of $\bbx$ into low, medium, and high frequency components which respectively represent signals that have slow, medium, and high variability with respect to the connectivity network between brain regions. This decomposition is utilized in this paper to analyze brain activity patterns associated with the learning of visual-motor tasks.

%
\section{Brain Signals during Learning}\label{sec_dataset}

We considered two experiments in which subjects learned a simple motor task 
\cite{bassett2013, Bassett2011}. In the experiments, fourty-seven right-handed participants (29 female, 18 male; mean age 24.13 years) volunteered with informed consent in accordance with the University of California, Santa Barbara Internal Review Board. After exclusions for task accuracy, incomplete scans, and abnormal MRI, 38 participants were retained for subsequent analysis. 

Twenty individuals participated in the first experimental framework. The experiment lasted 6 weeks, in which there were 4 scanning sessions, roughly at the start of the experiment, at the end of the 2nd week, at the end of the 4th week, and at the end of the experiment, respectively. During each scanning session, individuals performed a discrete sequence-production task in which they responded to sequentially presented stimuli with their dominant hand on a custom response box. Sequences were presented using a horizontal array of 5 square stimuli with the responses mapped from left to right such that the thumb corresponded to the leftmost stimulus. The next square in the sequence was highlighted immediately following each correct key press; the sequence was paused awaiting the depression of the appropriate key if an incorrect key was pressed. Each participant completed 6 different 10-element sequences. Each sequence consists of two squares per key. Participants performed the same sequences at home between each two adjacent scanning sessions, however, with different levels of exposure for different sequence types. Therefore, the number of trials completed by the participants after the end of each scanning session depends on the sequence type. There are 3 different sequence types (MIN, MOD, EXT) with 2 sequences per type. The number of trials of each sequence type completed after each scanning session averaged over the 20 participants is summarized in Fig. \ref{tab_trials}. During scanning sessions, each scan epoch involved 60 trials, 20 trials for each sequence type. Each scanning session contained a total of 300 trials (5 scan epochs) and a variable number of brain scans depending on how quickly the task was performed by the specific individual.

%
Eighteen subjects participated in the second experimental framework. The experiment had 3 scanning sessions spanning the three days. Each scanning session lasted roughly 2 hours and no training was performed at home between adjacent scanning sessions. Subjects responded to a visually cued sequence by generating responses using the four fingers of their nondominant hand on a custom response box. Visual cues were presented as a series of musical notes on a pseudo-musical staff with four lines such that the top line of the staff mapped to the leftmost key pressed with the pinkie finger. Each 12-note sequence randomly ordered contained three notes per line. Each training epoch involved 40 trials and lasted a total of 245 repetition times (TRs), with a TR of 2,000 ms. Each training session contained 6 scan epochs (240 trials) and lasted a total of 2,070 scan TRs.

In both experiments participants were instructed to respond promptly and accurately. Repetitions (e.g., ``11'') and regularities such as trills (e.g., ``121'') and runs (e.g., ``123'') were excluded in all sequences. The order and number of sequence trials were identical for all participants. Participants completed the tasks inside the MRI scanner for scanning sessions.

%
\begin{figure}[t]
\begin{center}{\footnotesize
\begin{tabular}{ c ccccc}        \toprule  
  & Session 1 & Session 2 & Session 3 & Session 4 \\\midrule
  MIN Sequences & $50$ & $110$ & $170$ & $230$ \\
  MOD Sequences & $50$ & $200$ & $350$ & $500$ \\
  EXT Sequences & $50$ & $740$ & $1430$ & $2120$ \\\bottomrule
\end{tabular}}
\caption{Relationship between training duration, intensity, and depth for the first experimental framework. The values in the table denote the number of trials (i.e., ``depth'') of each sequence type (i.e., ``intensity'') completed after each scanning session (i.e., ``duration'') averaged over the 20 participants.}
\label{tab_trials}
\end{center}
\end{figure}

%
\begin{figure}[t]
	\centering		
	\begin{minipage}[h]{0.24\textwidth}
		\centering
		\includegraphics[width=1 \textwidth]{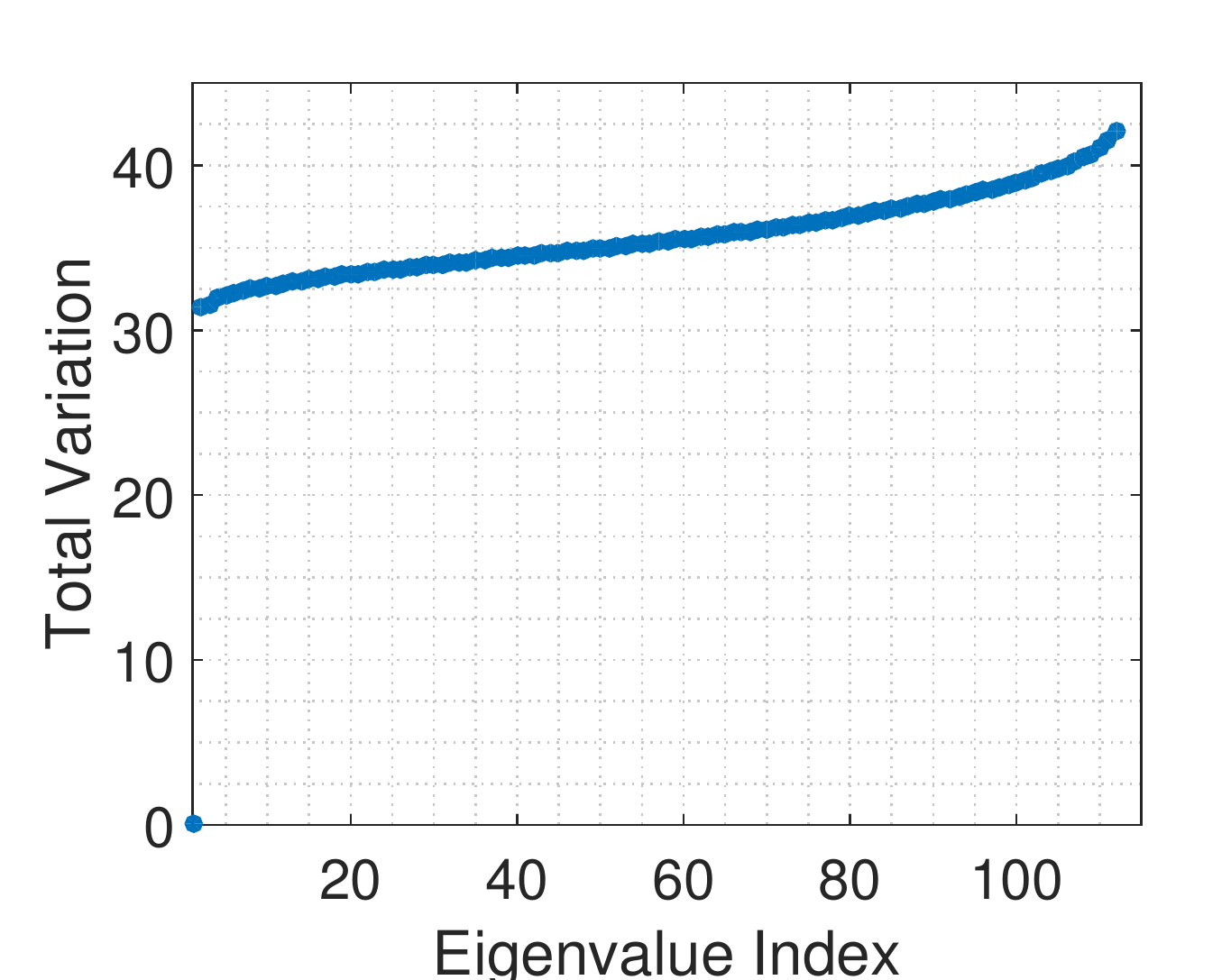}
		\scriptsize (a)
	\end{minipage}
	\begin{minipage}[h]{0.24\textwidth}
		\centering
		\includegraphics[width=1 \textwidth]{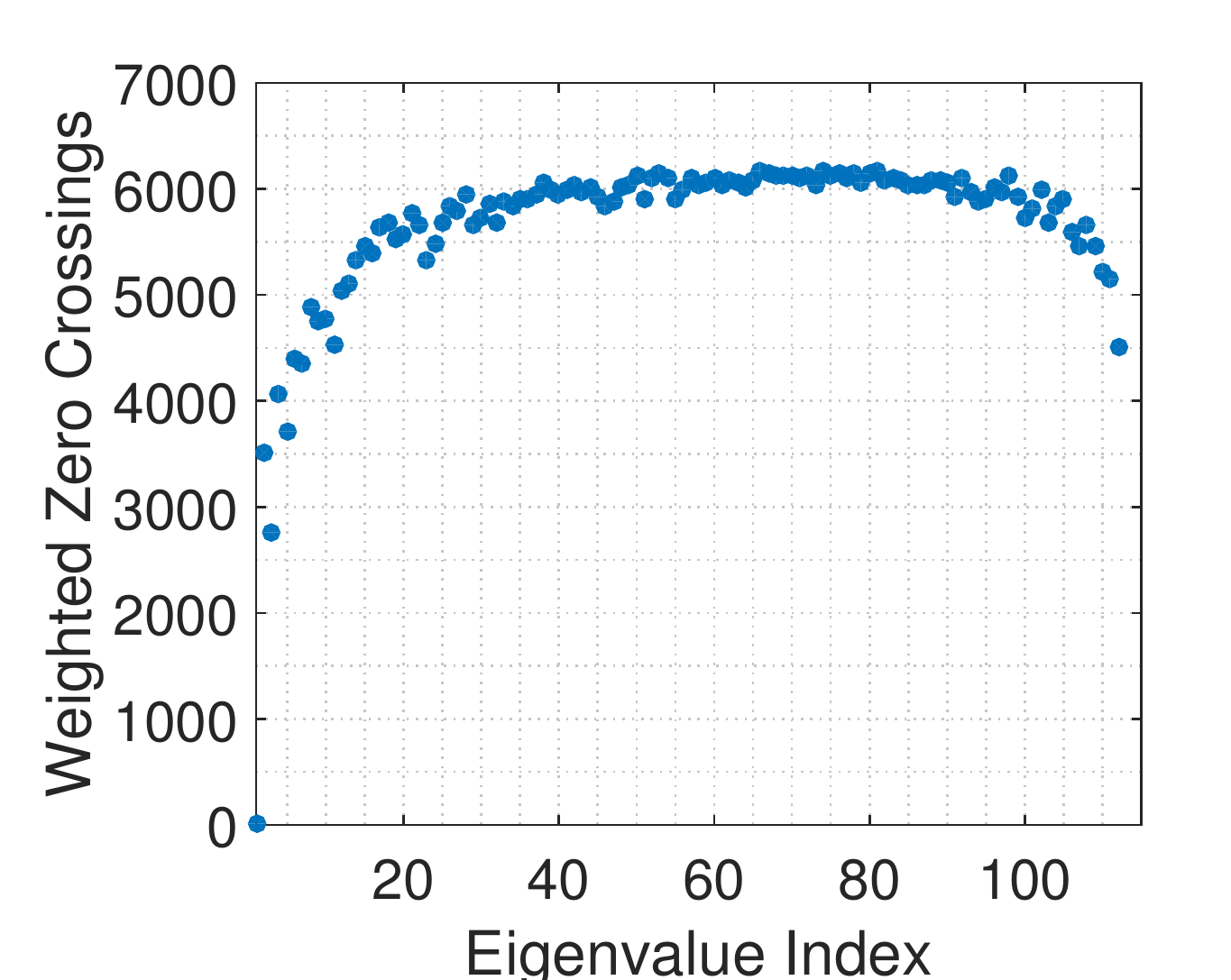}
		\scriptsize (b)
	\end{minipage}
	\begin{minipage}[h]{0.24\textwidth}
		\centering
		\includegraphics[width=1 \textwidth]{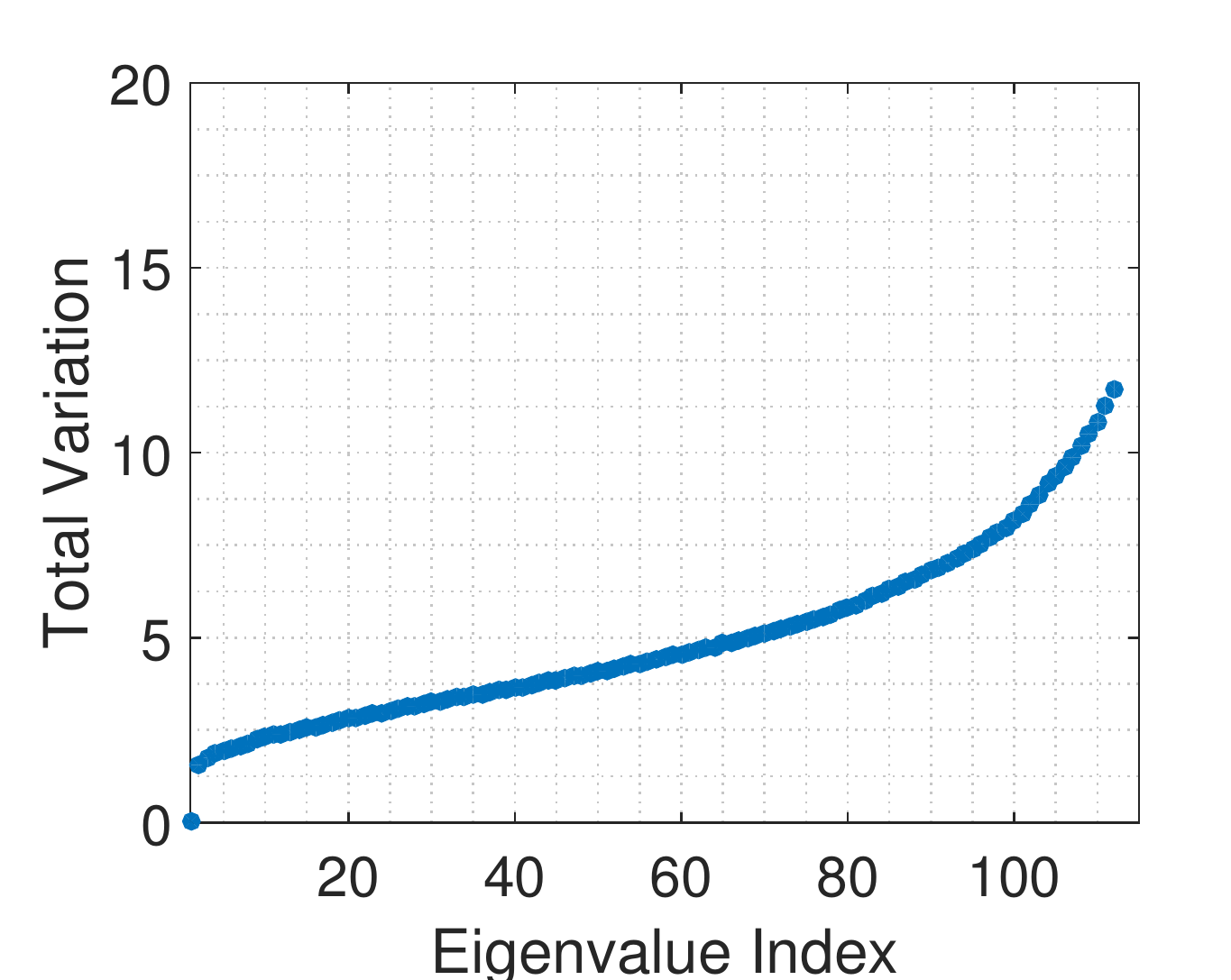}
		\scriptsize (c)
	\end{minipage}
	\begin{minipage}[h]{0.24\textwidth}
		\centering
		\includegraphics[width=1 \textwidth]{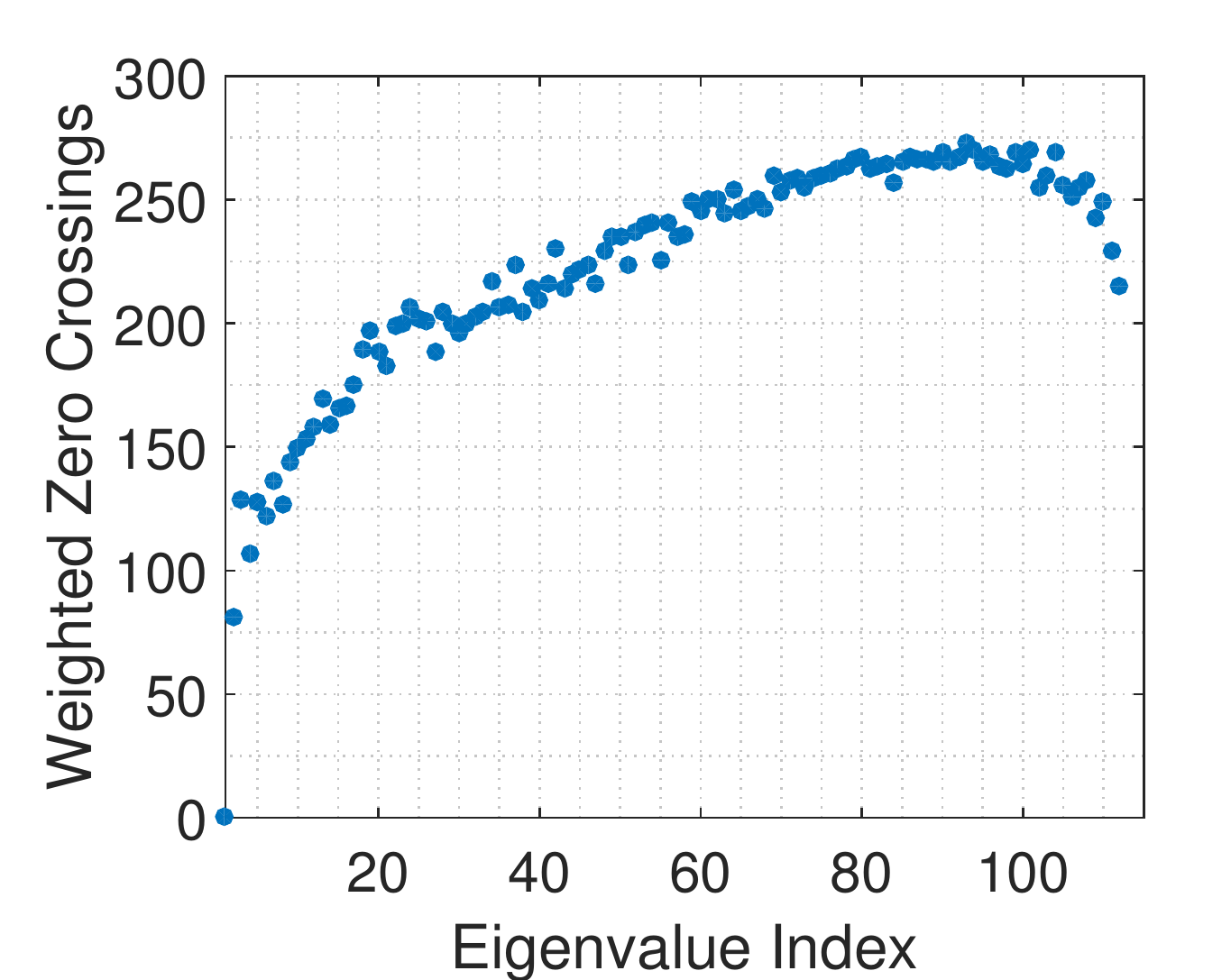}
		\scriptsize (d)
	\end{minipage}
	\caption{(a) Total variation $\TV(\bbv_k)$ and (b) weighted zero crossings $\ZC(\bbv_k)$ of the graph Laplacian eigenvectors for the brain networks averaged across participants in the 6 week training experiment. (c) and (d) present the values for the 3 day experiment. In both cases, the Laplacian eigenvectors associated with larger indexes vary more on the network and cross zero relatively more often, confirming the interpretation of the Laplacian eigenvalues as notions of frequencies. Besides, note that total variation increases relatively linearly with indexes. \vspace{-2mm}}
	\label{fig_TV}
\end{figure}

%
\begin{figure*}[t]
	\centering
	\includegraphics[trim=1cm 0cm 0cm 0cm, clip=true,width=0.98 \textwidth]{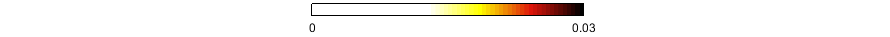}	
	\centering
	\includegraphics[width=0.98 \textwidth]{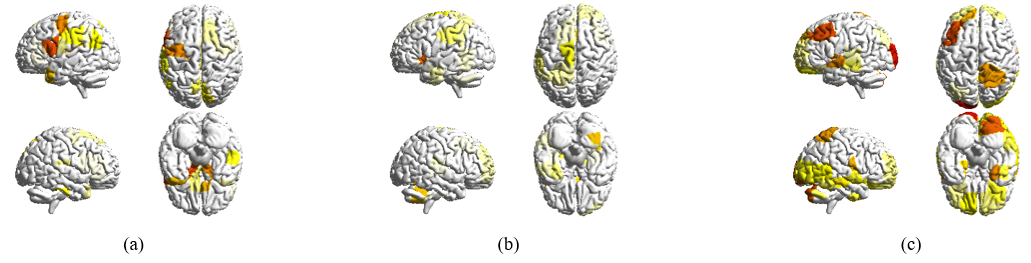}
	\centering
	\includegraphics[width=0.98 \textwidth]{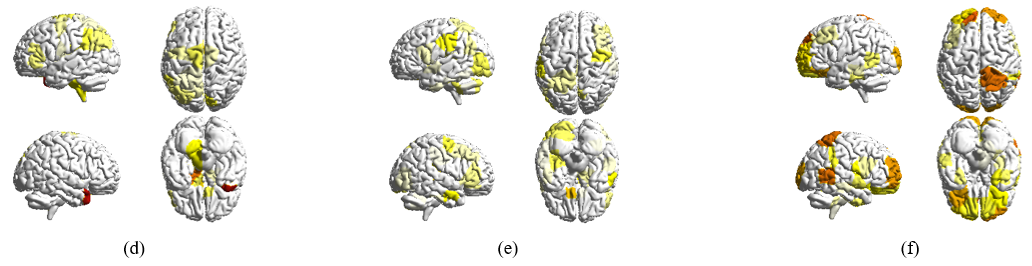}
	
	\caption{Absolute magnitude at each of the $n$ cortical structures averaged across participants in the 6 week experiment and averaged across all frequency components in (a) the set of low graph frequencies $\{ \bbv_k\}_{k=0}^{K_\L - 1}$, (b) the set of middle graph frequencies $\{ \bbv_k\}_{k=K_\L}^{K_\L + K_\M - 1}$, and (c) the set of high graph frequencies $\{\bbv_k\}_{k = K_\L + K_\M}^{n-1}$. (d)-(f) presents the average magnitudes for the 3 day experiment. Only brain regions with absolute magnitudes higher than a fixed threshold ($0.015$) are colored. Magnitudes across the datasets are highly similar in the low and high graph frequencies (correlation coefficients $0.5818$ and $0.6616$, respectively). The brain regions with high magnitude values significantly overlap with the visual and sensorimotor modules, in which more than $60 \%$ of values greater than the threshold belong to the visual and sensorimotor modules.}\vspace{-2mm}
	\label{fig_Network_Eig}
\end{figure*}

%
Reordering with fMRI was conducted using a 3.0 T Siemens Trio with a 12-channel phased-array head coil. For each functional run, a single-shot echo planar imaging sequence that is sensitive to blood oxygen level dependent (BOLD) contrast was utilized to obtain 37 (the first experiment) or 33 (the second experiment) slices (3mm thickness) per repetition time (TR), an echo time of 30 ms, a flip angle of 90$^{\circ}$, a field of view of 192 mm, and a 64 $\times$ 64 acquisition matrix. Image preprocessing was performed using the Oxford Center for Functional Magnetic Resonance Imaging of the Brain (FMRIB) Software Library (FSL), and motion correction was performed using FMRIB's linear image registration tool. The whole brain is parcellated into a set of $n = 112$ regions of interest that correspond to the 112 cortical and subcortical structures anatomically identified in FSL's Harvard-Oxford atlas. The choice of parcellation scheme is the topic of several studies in structural \cite{zalesky2010}, resting-state \cite{wang2009}, and task-based \cite{power2011} network architecture. The question of the most appropriate delineation of the brain into nodes of a network is open and is guided by the particular question one wants to ask. We use Harvard-Oxford atlas here because it is consistent with previous studies of task-based functional connectivity during learning \cite{Bassett2011, bassett2013}. The threshold in probability cutoff settings of Harvard Oxford atlas parcellation is $0$ so that no voxels were excluded.

For each individual fMRI dataset, we estimate regional mean BOLD time series by averaging voxel time series in each of the $n$ regions. We evaluate the magnitude squared spectral coherence \cite{sun2004} between the activity of all possible pairs of regions to construct $n\times n$ functional connectivity matrices $\bbW$. Besides, for each pair of brain regions $i$ and $j$, we use $t$-statistical testing to evaluate the probability $p_{i,j}$ of observing the measurements by random chance, when the actual data are uncorrelated \cite{he2007}. In the 3 day dataset, the value of all elements with no statistical significance ($p_{i,j} > 0.05$) \cite{Genovese2002threshold} are set to zero; the values remain unchanged otherwise. In the 3 day experiment, a single brain network is constructed for each participant. Thresholding is applied because the networks are for the entire span of the experiment and many entries in $\bbW$ would be close to zero without threshold correction. In the 6 week experiment, due to the long duration of the experiment, we build a different brain network per scanning session, per sequence type for each subject. Because each network describes the functional connectivity for one training session given a subject, not many entries will be removed even in the presence of threshold correction; consequently, no thresholding is applied for the 6 week dataset. We normalize the regional mean BOLD observations $\hat\bbx(t)$ at any sample time $t$ and consider $\bbx(t) = \hat\bbx(t) / \| \hat\bbx(t) \|_2$ such that the total energy of activities at all structures is consistent at different $t$ to avoid extreme spikes due to head motion or drift artifacts in fMRI.


%
\section{Brain Network Frequencies}\label{sec_network_illustration}

In this section, we analyze the graph spectrum brain networks of the dataset considered. For the brain network $\bbW$ of each subject, we construct its Laplacian $\bbL = \bbD - \bbW$, and evaluate the total variation $\TV(\bbv_k)$ [cf. \eqref{eqn_TV}] for each eigenvector $\bbv_k$. Fig. \ref{fig_TV} (a) and (c) plot the total variation of all graph eigenvectors averaged across participants of the 6 week training experiment and 3 day experiment, respectively. In both experiments, the Laplacian eigenvectors associated with larger indexes fluctuate more on the network. 
Another observation is that the total variation increases in linear scale with indexes.

Besides total variation, the number of zero crossings is used as a measure of the smoothness of signals with respect to an underlying network \cite{shuman2013}. Since brain networks are weighted, we adapt a slightly modified version -- weighted zero crossings -- to investigate the given graph eigenvector $\bbv_k$
\begin{align}\label{eqn_ZC}
    \ZC(\bbv_k) = \frac{1}{2} \sum_{i \neq j} w_{ij} \mathbb I \left\{ v_k(i) v_k(j) < 0\right\}.
\end{align}
In words, weighted zero crossings evaluate the weighted sum of the set of edges connecting a vertex with a positive signal to a vertex with a negative signal. Fig. \ref{fig_TV} (b) and (d) demonstrate the weighted zero crossings of all graph eigenvectors averaged across subjects of the 6 week and 3 day experiments, respectively. The weighted zero crossings increase proportionally with graph frequency index $k$ for $0 \leq k \leq 100$, as expected. However, eigenvectors associated with higher graph frequencies exhibit lower weighted zero crossings when $k$ is greater than $100$.

It would be interesting to examine where the associated eigenvectors lie anatomically, and the relative strength of their values. To facilitate the presentation, we consider three sets of eigenvectors, $\{ \bbv_k\}_{k=0}^{K_\L - 1}, \{ \bbv_k\}_{k=K_\L}^{K_\L + K_\M - 1}$ and $\{\bbv_k\}_{k = K_\L + K_\M}^{n-1}$, and compute the absolute magnitude at each of the $n$ cortical and subcortical regions averaged across participants and across all graph frequencies belonging to each of the three sets. Fig. \ref{fig_Network_Eig} presents the average magnitudes for the two experimental frameworks considered in the paper using BrainNet \cite{xia2013}, where brain regions with absolute magnitudes lower than a fixed threshold are not colored. Throughout the paper, the parameter $K_\L$ is set as $40$ and $K_\M$ is set as $32$. We chose this combination because it yields three roughly equally-sized components with one piece corresponding to the $40$ lowest graph frequencies and another piece corresponding to the $40$ highest frequencies. The results presented in the paper are robust with the choice of parameters: we examined the results for $K_\L$ and $K_\M$ in the range of $32$ to $42$ and found similar observations as the ones presented. To demonstrate that, we also consider the parameters $K_\L = 38$, $K_\M = 37$ and $K_\L = 35$, $K_\M = 42$. The average correlation coefficient between the absolute magnitude across brain regions of the parameters selected for this paper and the two chosen for robustness is $0.8659$. See Fig. \ref{fig_corr_R} for another quantification on the robustness of parameters. In this paper, we report correlation coefficient as a quantification of similarity measure when we examine the similarity between two vectors. We also investigate cosine similarity and find high similarities as well.

%
\subsection{Artificial Functional Brain Networks}\label{sec_artificial_network}

An approach to analyze the complex networks is to define a model to generate artificial networks  \cite{lohse2014}. The main motivation of an artificial network model is to use them to analyze complex brain networks. Examples of such models include Erd\H{o}s-R\'{e}nyi model \cite{erdds1959} for unweighted networks, Barab\'{a}si-Albert model for scale free network \cite{albert2002}, and recent development and insights on weighted network models \cite{lohse2014}. Here we present a framework to construct artificial networks that can be used to mimic the functional brain networks with only a few parameter inputs. The model is related to weighted block stochastic model \cite{aicher2014}, but involves more aspects like individual variance and links independent of their connecting regions. The output of the method would be a symmetric network with edge weights between $0$ and $1$ without self-loops. 

To begin, suppose the desired network has two clusters of nodes $\ccalV_1$ and $\ccalV_o$. The algorithm requires the average edge weight $\mu_1$ for connections between nodes of the first cluster $\ccalV_1$, average edge weight $\mu_o$ for links between nodes of the other cluster $\ccalV_o$, and average edge weight $\mu_{1o}$ for inter-cluster connections. To reflect the fact that the edge weights on some links are independent of their joining vertices, for each edge within $\ccalV_1$, with probability $p_\epsilon < 1$, its weight is randomly generated with respect to uniform distribution $\ccalU[0, 1]$ between $0$ and $1$, and with probability $1 - p_\epsilon$, its weight is randomly generated with respect to uniform distribution $\ccalU[\mu_1 - \delta, \mu_1 + \delta]$. The parameter $p_\epsilon$ determines the percentage of edges whose weights are selected irrespective of their actual locations To further simulate the observation that different participants may possess distinctive brain networks, if the edge weight is randomly generated from a uniform distribution $\ccalU[\mu_1 - \delta, \mu_1 + \delta]$, it is then perturbed by $w_u \sim \ccalU[-u_\epsilon / 2, u_\epsilon/2]$ where $u_\epsilon$ controls the level of perturbation. The edge weights for connections within cluster $\ccalU_o$ are generated similarly: with probability $1 - p_\epsilon$, the edge weight is randomly chosen from the uniform distribution $\ccalU[\mu_o - \delta, \mu_o + \delta]$ before being contaminated by $w_u \sim \ccalU[-u_\epsilon / 2, u_\epsilon/2]$. The edge weights for connections between clusters $\ccalU_1$ and $\ccalU_o$ are formed analogously using $\mu_{1o}$. The method presented here can be easily generalized to analyze brain networks with more regions of interest, i.e. by specifying sets of regions of interest and by detailing the expected correlation values on each type of connection between different regions.

%
\begin{remark}\label{remark_artificial_network}\normalfont
At one extreme we can make each node $i$ belonging to a different set $\ccalV_i = \{i\}$. Then the method requires the inputs of expected weights for all nodes, or alternatively speaking, the expected network. At the other extreme, there is only one set of nodes $\ccalV$, and then the method is highly akin to a network with edge weights completely randomly generated. Any construction of interest would have some prior knowledge regarding the community structure. Therefore, the method proposed here can be used to see if the network constructed with the specific choice of community structure highly simulate the key properties of the actual network, and can be used to examine the evolution of community structure in the brain throughout the process to master a particular task.
\end{remark}\vspace{-2mm}

%
\begin{figure*}[t]
\centering
\includegraphics[trim=0cm 0cm 0cm .2cm, clip=true, width=0.95 \textwidth]{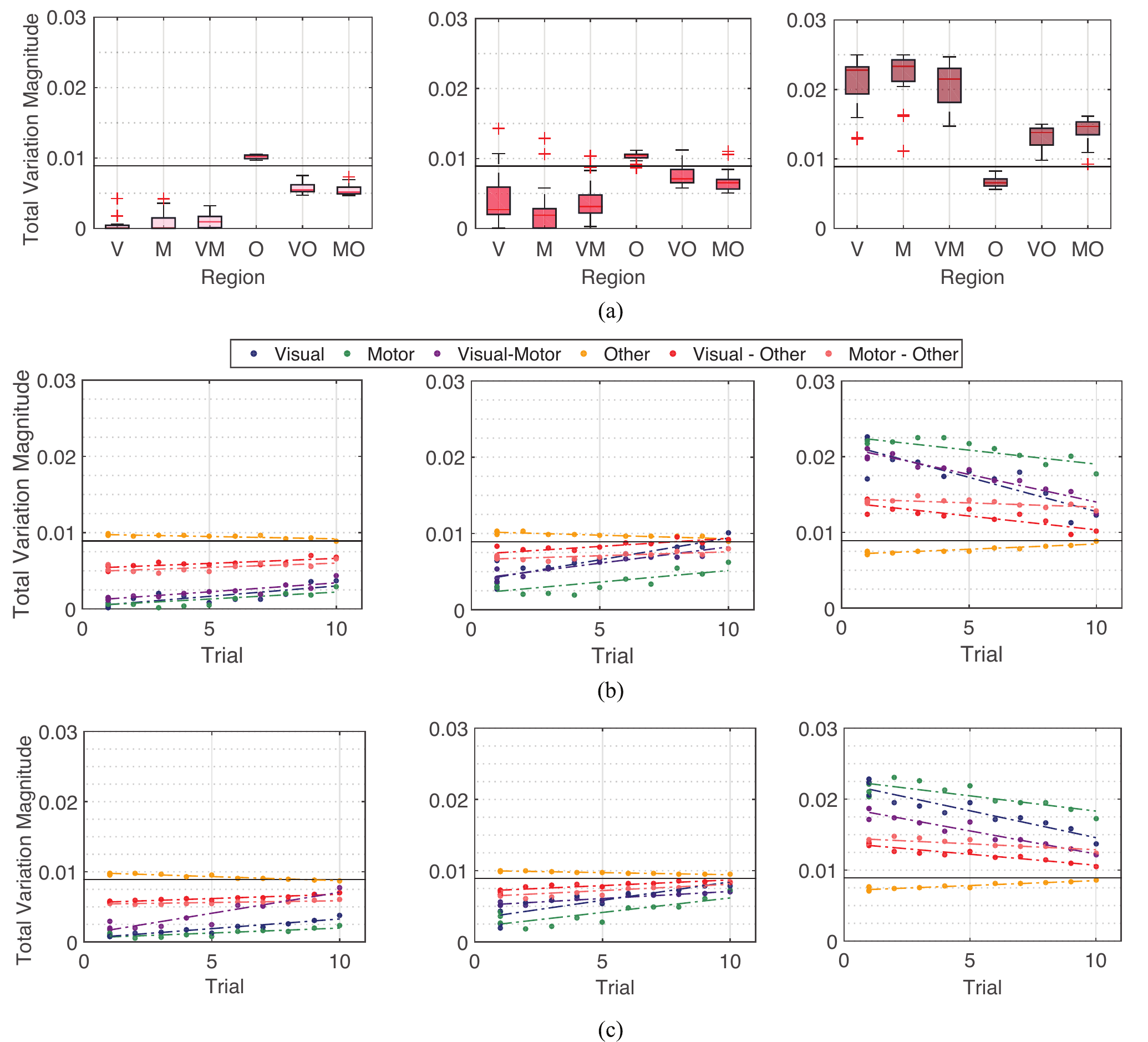}
\caption{Spectral property of brain networks in the 6 week experiment. (a) Left: Averaged total variation of eigenvectors $\bbv_k$ for 6 different types of connections of the brain averaged across all eigenvectors associated with low graph frequencies $\bbv_k \in \{\bbv_k\}_{k=0}^{K_\L - 1}$, across all participants and scan sessions. Middle: Across all eigenvectors with middle graph frequencies $\bbv_k \in \{\bbv_k\}_{k = K_\L}^{K_\L  + K_\M - 1}$. Right: Across all eigenvectors with high graph frequencies $\bbv_k \in \{\bbv_k\}_{k = K_\L + K_\M}^{n-1}$. (b) Median total variations of brain networks across participants of different scanning sessions and different sequence types with respect to the level of exposure of participants to the sequence type at the scanning session. Relationship between training duration, intensity, and depth is summarized in Fig. \ref{tab_trials}. Value of $1$ on the $x$-axis in the figures refers to minimum exposure to sequences (all 3 sequence types of the first session), and value of $10$ on the $x$-axis denotes the maximum exposure to sequences (EXT sequence types of the fourth session). An association between spectral property of brain networks and the level of exposure is clearly observed (average correlation coefficient $0.8164$). (c) Median total variations evaluated upon artificial networks. Spectral properties of actual brain networks can be closely simulated using a few parameters. The main text gives all correlation values for similarity between variance among subjects and between correlations of training intensity. \vspace{-2mm}}
\label{fig_TV_Mag_6Week}
\end{figure*}

%
\subsection{Spectral Properties of Brain Networks}\label{sec_spectral_property}

In this section, we analyze graph spectral properties of brain networks. Given the graph Laplacian, we examine the fluctuation of its eigenvectors on different types of connections in the brain network \cite{bassett2015}. More specifically, given an eigenvector $\bbv_k$, its variation on the visual module is defined as
\begin{align}\label{eqn_TV_visual}
    \text{TV}_\text{visual}(\bbv_k) = \frac{\sum_{i, j \in \ccalV_\text v, i \neq j} w_{ij} (v_k(i) - v_k(j))^2 }
        {\sum_{i, j \in \ccalV_\text v, i \neq j} w_{ij}},
\end{align}
where $\ccalV_\text v$ denotes the set of nodes belonging to the visual module. The measure $\TV_\visual(\bbv_k)$ computes the difference for signals on the visual module for each unit of edge weight. To facilitate interpretation, we only consider three sets of eigenvectors $\{ \bbv_k\}_{k=0}^{K_\L - 1}, \{ \bbv_k\}_{k=K_\L}^{K_\L + K_\M - 1}$, and $\{ \bbv_k\}_{k=K_\L + K_\M}^{n-1}$. We then compute the visual module total variation $\TV_\visual^\L$ averaged over eigenvectors $\{ \bbv_k\}_{k=0}^{K_\L - 1}$, and $\TV_\visual^\M$ as well as $\TV_\visual^\HH$ similarly. Besides $\text{TV}_\text{visual}$, we also examine the level of fluctuation of eigenvectors on edges within the motor module, denoted as $\TV_\motor$, and on connections belonging to brain modules other than the visual and motor module $\TV_\other$. Further, there are links between two separate brain modules, and to assess the variation of eigenvectors on those links, we define total variations between the visual and motor modules 
\begin{align}\label{eqn_TV_visual_motor}
    \text{TV}_\vm(\bbv_k) = \frac{\sum_{i \in \ccalV_\text v, j \in \ccalV_\text m} w_{ij} (v_k(i) - v_k(j))^2 }
        {\sum_{i \in \ccalV_\text v, j \in \ccalV_\text m} w_{ij}},
\end{align}
where $\ccalV_\text m$ denotes the set of nodes belonging to the motor module. Total variations $\TV_\vo$ between the visual and other modules, and total variations $\TV_\mo$ between the motor and other modules are defined analogously. We chose to study visual and motor modules separately from other brain modules because of their well-known associations with motor learning \cite{kleim2002, bassett2015}. 

Fig. \ref{fig_TV_Mag_6Week} (a) presents boxplots of the variation for eigenvectors of different graph frequencies measured over different types of connections across participants, at the start of the six week training. Despite that total variation of eigenvectors should increase with their frequencies, the variation on the other module $\TV_\other^\L$ of eigenvectors associated with low frequencies are higher than $\TV_\other^\HH$ (pass t-test with $p < 0.0001$). This observation is discussed in detail in Section \ref{sec_network_discussion}.

%
\begin{figure*}[t]		
        \centering
        \includegraphics[trim=1cm 0cm 0cm 0cm, clip=true,width=0.98 \textwidth]{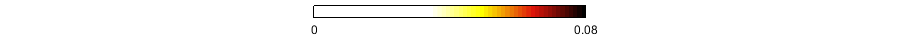}
		
        \centering
        \includegraphics[width=0.98 \textwidth]{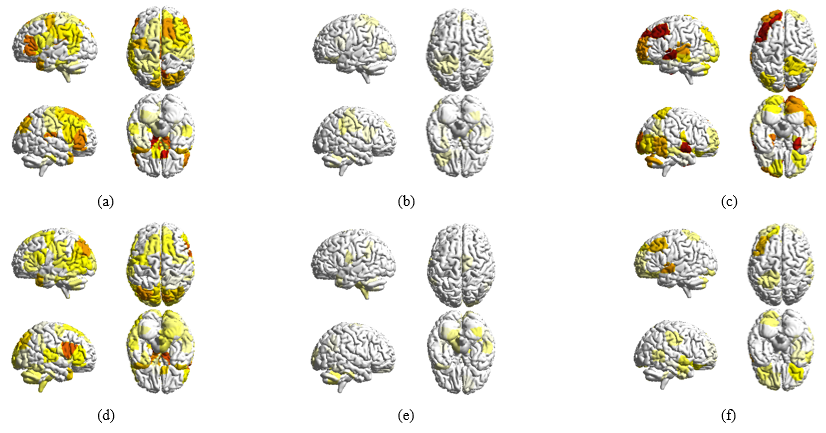}
		
        \caption{Distribution of decomposed signals for the 6 week experiment. (a) Absolute magnitudes for all brain regions with respect to $\bbx_\L$ -- brain signals varing smoothly across the network -- averaged across all sample points for each individual and across all participants at the first scan session of the 6 week dataset. (b) With respect to $\bbx_\M$ and (c) with respect to $\bbx_\HH$ -- signals rapidly fluctuating across the brain. (d), (e), and (f) are averaged $\bbx_\L, \bbx_\M$ and $\bbx_\HH$ at the last scan session of the 6 week dataset, respectively. Only regions with absolute magnitudes higher than a fixed threshold are colored.}
        \label{fig_Signal_6Week}
\end{figure*}

\begin{figure*}[t]		
        \centering
        \includegraphics[trim=1cm 0cm 0cm 0cm, clip=true,width=0.98 \textwidth]{Colorbar_sig.png}
	
        \centering
        \includegraphics[width=0.98 \textwidth]{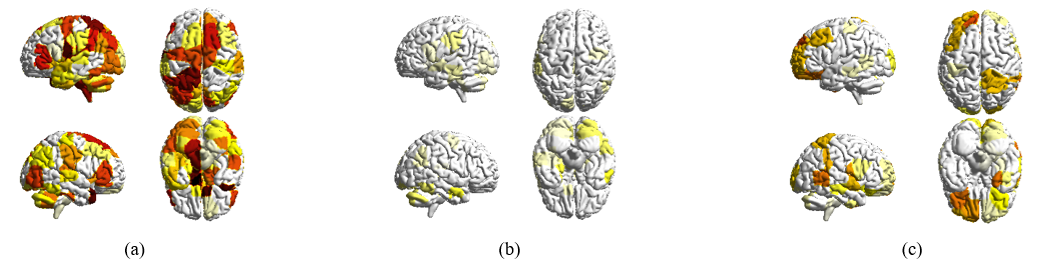}
        \caption{Distribution of decomposed signals for the 3 day experiment. (a), (b), and (c) are $\bbx_\L, \bbx_\M$ and $\bbx_\HH$ averaged across all sample points for each subject and across participants in the 3 day experiment, respectively. Regions with absolute value less than a threshold are not colored. }
        \label{fig_Signal_3Day}
\end{figure*}

%
Next we study how the graph spectral properties of brain networks evolve as participants become more familiar with the tasks. Fig. \ref{fig_TV_Mag_6Week} (b) illustrates the median of the variation for eigenvectors of different graph frequencies measured over different types of connections across subjects, at 10 different levels of exposure in the six week training. As participants become more acquainted with the assignment, their brain networks display lower variation in the visual and motor modules and higher variation in the other modules for low and middle graph frequencies, and the exact opposite is true for high graph frequencies. The association with training intensity is statistically significant (average correlation coefficient $r = 0.8164$). 

%
\begin{figure*}[t]
	\centering
	\includegraphics[trim=0cm 0cm 0cm 0cm, clip=true,width=0.9 \textwidth]{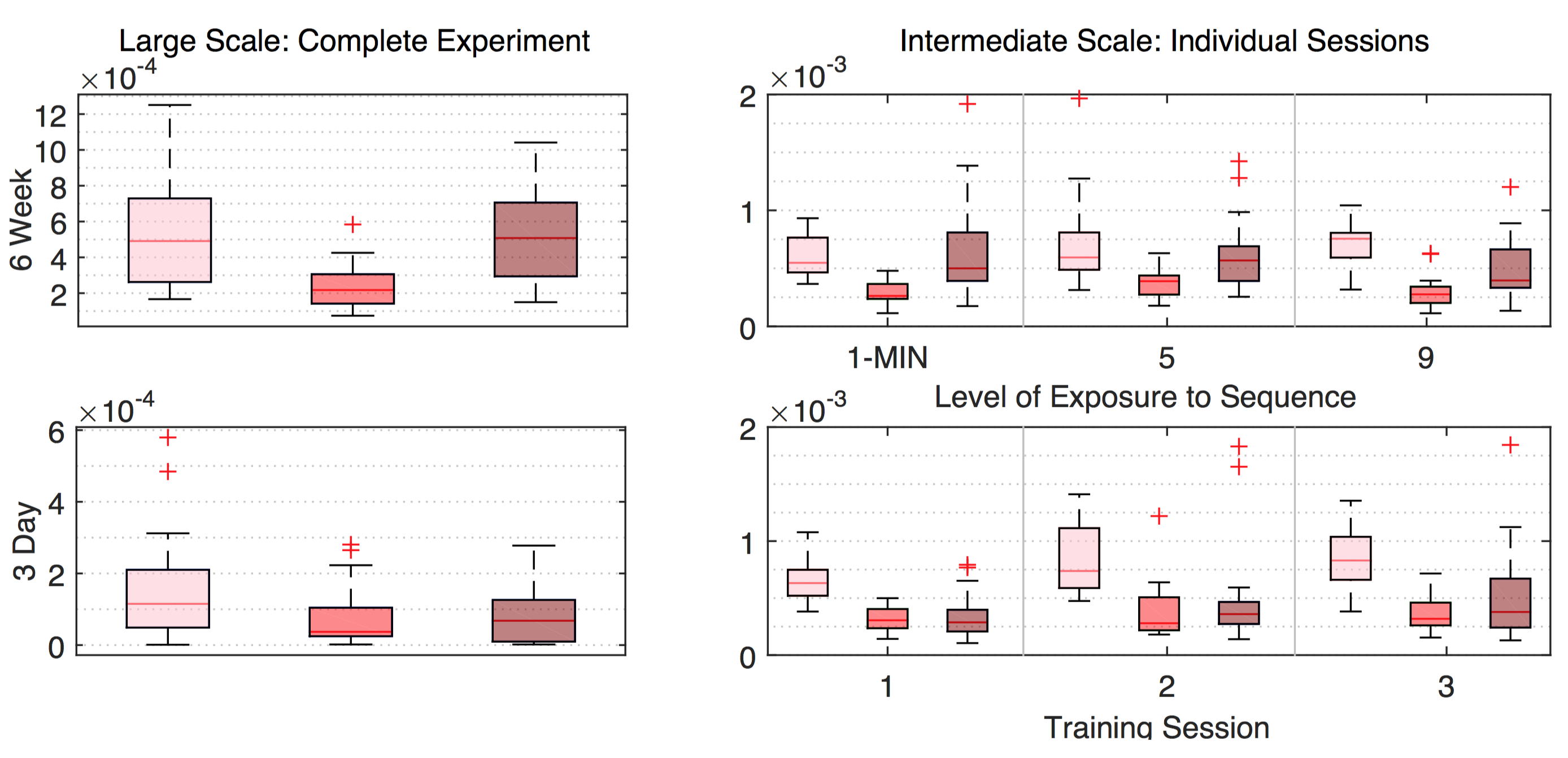}
	\caption{Temporal adaptations of spatial variations. Boxplots showing differences in temporal adaptabilities between brain activities with smooth (pink), moderate (red) and rapid (maroon) spatial variations, measured over the complete experiment (Left) and individual training sessions (Right) for 6 week experiment (Top) and 3 day experiment (Bottom). We measured the temporal adaptations using the variance of the averaged activities over the complete experiment or with individual training sessions. Compared to activities with moderate spatial variations, smooth ($95\%$ sessions pass t-test with $p < 0.01$) and rapid ($65\%$ sessions pass t-test with $p < 0.005$) spatial variations have significantly higher temporal adaptations.}
	\label{fig_GFT_Mag_Sub}
\end{figure*}

%
\subsection{Discussion}\label{sec_network_discussion}

Firstly, we examine why we see a decrease in zero crossings of graph frequencies when $k$ is greater than $100$ in Fig. \ref{fig_TV}. A detailed analysis shows this is because the functional brain networks are highly connected with nearly homogeneous degree distribution, and consequently each high graph frequency tends to have a value with high magnitude at one vertex of high degree and similar values at other nodes, resulting in a smaller global zero crossings for eigenvectors associated with very high frequencies.

Secondly, in terms of the visualization of graph frequencies in Fig. \ref{fig_Network_Eig}, the most interesting finding relates to the eigenvectors associated with high graph frequencies. The magnitudes at different brain regions for high frequencies are significantly similar across the two datasets investigated (correlation coefficient $0.6616$), and brain regions with high magnitude values are highly alike (greater than $60\%$ overlap) to the visual and sensorimotor cortices \cite{braitenberg2013}. This is likely to be a consequence of the fact that visual and motor regions are more strongly connected with other structures, and hence an eigenvector with a high magnitude on visual or motor structures would result in high global spatial variation. The eigenvectors of low graph frequencies are more spread across the networks, resulting in low global variations. The middle graph frequencies are less interesting -- the magnitudes at most regions (greater than $90\%$) do not pass the threshold, and little associations (correlation coefficient $0.3529$) can be found between the eigenvectors of the 6 week and 3 day experiments. 

Thirdly, to better interpret the meaning of variations for specific types of connections, we construct artificial networks as described in Section \ref{sec_artificial_network} with visual and motor modules as regions of interest, and consider other modules to be brain regions other than visual and motor modules. We observe that there are three contributing factors that cause the variation within a specific module to become higher for higher eigenvectors and to become lower for lower eigenvectors: (i) Increases in the average edge weight for connections within the module, (ii) Increments in the average edge weight for links between this module to other module, and (iii) Escalation in the average edge weight for associations within the other module. This can also be observed by analyzing closely the definition of total variation. If a module is highly connected, in order for the eigenvector associated with a low graph frequency to be smooth on the entire network, it has to be smooth on the specific module, resulting in a low value in the variation of an eigenvector associated with a low graph frequency with respect to the module of interest. Similarly, the increase in the variation of connections between two modules, e.g. between visual and other modules are resulted from: (i) The growth in the average edge weight for connections between visual and other modules, or (ii) The augmentation of average weight for links within the other module. 
The graph spectral properties as in Fig. \ref{fig_TV_Mag_6Week} (a) are observed because (i) visual and motor modules are themselves highly connected, and (ii) visual module is also strongly linked with motor module. 

Finally, in analyzing the evolution of graph spectral properties as participants become more familiar with the tasks, following the interpretations based on artificial network analysis, this evolution in graph spectral properties of brain networks is mainly caused by the decrease in values of connections within visual and motor modules and between the visual and motor modules. An interesting observation is that the values in the variation of eigenvectors associated with high frequencies decline with respect to the visual module much faster than that of motor module, even though the visual module is more strongly connected throughout training compared to the motor module. A deep analysis using artificial networks shows that this results from the following three factors: (i) Though more strongly connected compared to the motor module, connections within the visual module weaken very quickly, (ii) The motor module is more closely connected with the other module than the link between the visual module to the other module, and (iii) Association levels within the other module stay relatively constant. Therefore, as participants become more exposed to the tasks, compared to the visual module, the motor module becomes more strongly connected. 
 The graph spectral properties of actual brain networks and their evolution can be closely imitated using artificial networks as plotted in Fig. \ref{fig_TV_Mag_6Week} (c). The artificial network created for our analysis best imitated the real brain networks with parameters $p_\epsilon$ of $0.10$, $u_\epsilon$ of $0.10$, and $\delta$ of $0.01$. The average edge weights $\mu$ for visual ($v$), motor ($m$), other ($o$), and inter-connecting regions are $\mu_v = 0.6028$, $\mu_m = 0.4902$, $\mu_o = 0.3098$, $\mu_{vm} = 0.3985$, $\mu_{vo} = 0.3181$, and $\mu_{mo} = 0.3271$. The correlation coefficients of association with training intensity between real and artificial networks for low, medium, and high graph frequencies are $0.6436$, $0.7187$ and $0.8457$, respectively. Additionally, the variation among participants in real dataset can be highly mimicked using artificial network model we proposed, with correlation coefficients $0.9338$, $0.9660$, and $0.9486$ for low, medium, and high graph frequencies, respectively. The analysis for the three day training dataset is highly similar (correlation coefficients $0.9834$, $0.9186$, and $0.9674$ for low, medium, and high graph frequencies, respectively) and for this reason we do not present and analyze it separately here.

%
\section{Frequency Decomposition of Brain Signals}\label{sec_signal_decomp}

The previous sections focus on the study of brain networks and their graph spectral properties. In this section, we investigate brain signals from a GSP perspective, and analyze the brain signals by examining the decomposed graph signals $\bbx_\L, \bbx_\M$, and $\bbx_\HH$ with respect to the underlying brain networks. We compute the absolute magnitude of the decomposed signal $\bbx_\L$ for each brain region averaged across all sample signals for each individual during a scan session and then averaged across all participants. Similar aggregation is applied for $\bbx_\M$ and $\bbx_\HH$.

Fig. \ref{fig_Signal_6Week} presents the distribution of the decomposed signals corresponding to different levels of spatial variations for the first scan session (top row) and the last scan session (bottom row) in the 6 week experiment. Fig. \ref{fig_Signal_3Day} exhibits how the decomposed signals are distributed across brain regions in the 3 day experiment. Brain regions with absolute magnitudes lower than a fixed threshold are not colored. 

%
\subsection{Temporal Variation of Graph Frequency Components}\label{sec_variance}

We analyze temporal variation of decomposed signals with respect to different levels of spatial variations. To that end, we evaluate the variance of the decomposed signals over multiple temporal scales -- over days and minutes \cite{Newell2009} -- for the two experiments. We describe the method specifically for $\bbx_\L$ for simplicity and similar computations were conducted for $\bbx_\M$ and $\bbx_\HH$. At the macro timescale, we average the decomposed signals $\bbx_\L$ for all sample points within each scanning session with different sequence type, and evaluate the variance of the magnitudes of the signals \cite{garrett2012} across all the scanning sessions and sequence types. For the 6 week experiment, there are 4 scanning sessions and 3 different sequence types, so the variance is with respect to 12 points. For the 3 day experiment, there are 3 scanning sessions and only 1 sequence type, and therefore the variance is for 3 points. As for the micro or minute-scale, we average the decomposed signals $\bbx_\L$ for all sample points within each minute, and evaluate the variance of the magnitudes of the averaged signals across all minute windows for each scanning session with different sequence types. The evaluated variance is then averaged across all participants of the experiment of interest. 

Fig. \ref{fig_GFT_Mag_Sub} displays the variance of the decomposed signals $\bbx_\L, \bbx_\M$ and $\bbx_\HH$ at two different temporal scales of the two experiments. For the 6 week dataset, 3 session-sequence combinations, with the number proportional to the level of exposure of participants to the sequence ($1$-MIN refers to MIN sequence at session 1, $5$ denotes MIN sequence at session 4, $9$ entails EXT sequence at session 3) are selected out of the 12 combinations in total for a cleaner illustration, but all the other session-sequence combinations exhibit similar properties.

%
\subsection{Discussion}\label{sec_variance_discussion}

A deep analysis of Figs. \ref{fig_Signal_6Week} and \ref{fig_Signal_3Day} yields many interesting aspects of graph frequency decomposition. First, for $\bbx_\L$, the magnitudes on adjacent brain regions tend to possess highly similar values, resulting in a more evenly spread brain signal distribution, where as for $\bbx_\HH$, neighboring signals can exhibit highly dissimilar values; this corroborates the motivation to use graph frequency decomposition to segment brain signals into pieces corresponding to different levels of spatial fluctuations. Second, decomposed signal of a specific level of variations, e.g. $\bbx_\L$ and $\bbx_\HH$, are highly similar with respect to different scan sessions in an experiment as well as with respect to two experiments with different sets of participants. The correlation coefficient between datasets for low, medium and high graph frequencies are $0.5841$, $0.2852$, and $0.6469$, respectively. This reflects the fact that frequency decomposition is formed by applying graph filters with different pass bands upon signals and therefore should express some consistent aspects of brain signals. Third, recall that we normalize the brain signals at every sample point for all subjects, and for this reason signals $\bbx_\L, \bbx_\M$ and $\bbx_\HH$ would be similarly distributed across the brain if nothing interesting happens at the decomposition. However, in both Fig.s \ref{fig_Signal_6Week} and \ref{fig_Signal_3Day}, it is observed that many brain regions possess magnitudes higher than a threshold in $\bbx_\L$ ($\sim60\%$ pass) and $\bbx_\HH$ ($\sim20\%$ pass) while not many brain regions pass the thresholding with respect to $\bbx_\M$ ($\sim3\%$ pass). It has long been understood that the brain combines some degree of disorganized behavior with some degree of regularity and that the complexity of a system is high when order and disorder coexist \cite{Sporns2011}. $\bbx_\L$ varies smoothly across the brain network and therefore can be regarded as regularity (order), whereas $\bbx_\HH$ fluctuates rapidly and consequently can be considered as randomness (disorder). This evokes the intuition that graph frequency decomposition segments a brain signal $\bbx$ into pieces $\bbx_\L$ and $\bbx_\HH$, which reflect order and disorder (and are therefore more interesting), as well as the remaining $\bbx_\M$.

For the variance analysis, it is expected for the low graph frequency components (smooth spatial variation) to exhibit the smallest temporal variations, exceeded by medium and then high counterparts. Nonetheless, it is observed that brain activities with smooth spatial variations exhibit the most rapid temporal variation. Because it has been shown that temporal variation of observed brain activities is associated with better performance in tasks \cite{garrett2012}, this indicates a stronger contribution of low graph frequency components during the learning process. Furthermore, since the measurements were normalized such that the total energy of overall brain activities stayed constant at different sampling points, the rapid temporal changes of low graph frequency components should be accompanied by fast temporal variation of some other components, which are found to be high frequency components in all cases. Because these results were consistent for all of the temporal scales and datasets that we examined, and the association between temporal variability and positive performance has been established \cite{Heisz2012}, we concluded that brain activities with smooth or rapid spatial variations offer greater contributions during learning. The graph frequency signatures at different stages of learning is analyzed in the next section.

%
\section{Frequency Signatures of Task Familiarity}\label{sec_learning}

%
\begin{figure}[t]
	\begin{center}{\footnotesize
			\begin{tabular}{ c ccc}        \toprule  
				& $\|\bbx_\L\|_2$ & $\|\bbx_\M\|_2$ & $\|\bbx_\HH\|_2$ \\\midrule
				6 week experiment (linear scale) & $-0.3155$ & $0.0897$ & $0.4125$ \\
				6 week experiment (logarithm scale) & $-0.5409$ & $0.3992$ & $0.3565$\\
				3 day experiment & $-0.9873$ & $0.8443$ & $0.9605$ \\\bottomrule
			\end{tabular}}
			\caption{Pearson correlation coefficients between the number of trials (level of task familiarity) and R values, defined as correlations between learning rate parameters and the norm of the decomposed signal of interest. More obvious adaptability between decomposed signals and learning across training is observed for $\bbx_\L$ and $\bbx_\HH$, with decreasing association with exposure to tasks for the former and increasing importance for the latter.}
			\label{fig_corr_R}
		\end{center}
	\end{figure}

Given that the decomposed signals exhibit interesting perspectives, it is natural to probe whether the signals corresponding to different levels of spatial variations associate with learning. To that end, we first describe how learning rate is evaluated. Given a participant, for each sequence completed, we defined the movement time $M$ as the difference between the time of the first button press and the time of the last button press during a single sequence. We then estimate the participant's learning rate by fitting an exponential function (plus a constant) using the robust outlier correction \cite{rosenbaum2009} to the sequence of movement times $\bbM$
\begin{align}\label{eqn_exponential_fit}
    \bbM = c_1 e^{\bbt / \kappa} + c_2.
\end{align}
where $\bbt$ is a sequence representing the time index, $\kappa$ is the exponential drop-off parameter (which we call the ``learning rate parameter'') used to describe the early and fast rate of improvement, and $c_1$ and $c_2$ are nonnegative constants. Their sum $c_1 + c_2$ is an estimation of the starting speed of the participant of interest prior to training, while the parameter $c_2$ entails the fastest speed to complete the sequence attained by that participant after extended training. A negative value of $\kappa$ indicates a decrease in movement time $M(t)$, which is thought to indicate that learning is occurring \cite{dayan2011}. 
We chose exponential because it is viewed as the most statistically robust choice \cite{heathcote2000}. Further, the approach that we used has the advantage of estimating the rate of learning independent of initial performance or performance ceiling.

%
\begin{figure*}[t]
	\centering
	\includegraphics[width=0.95 \textwidth]{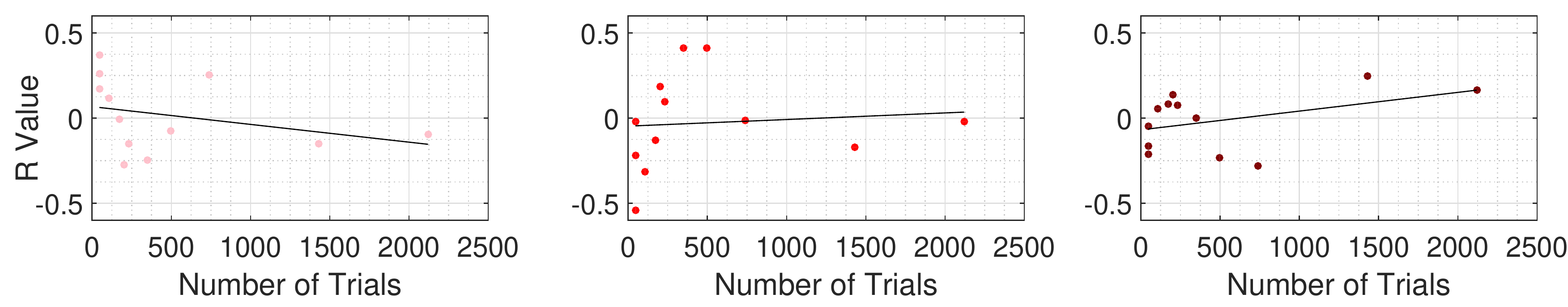}
	
	\includegraphics[width=0.95 \textwidth]{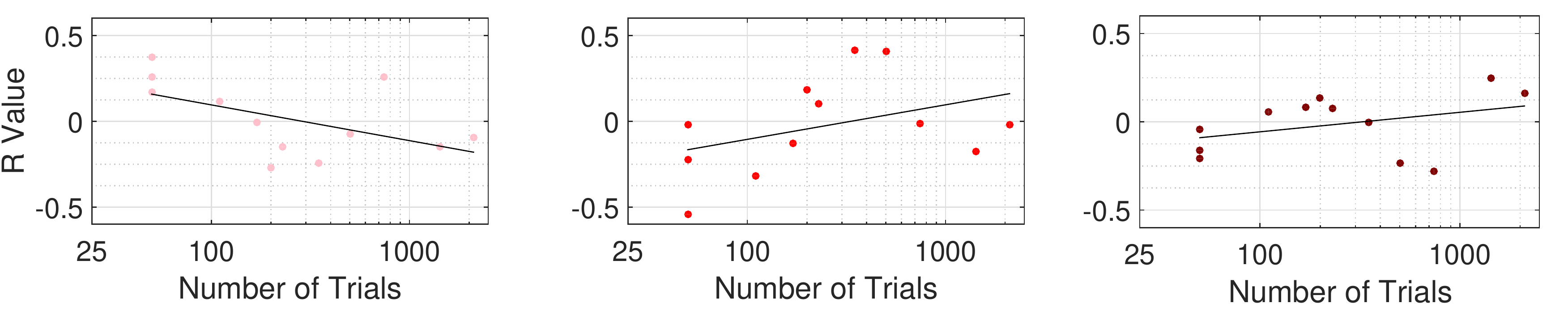}
	
	\includegraphics[width=0.95 \textwidth]{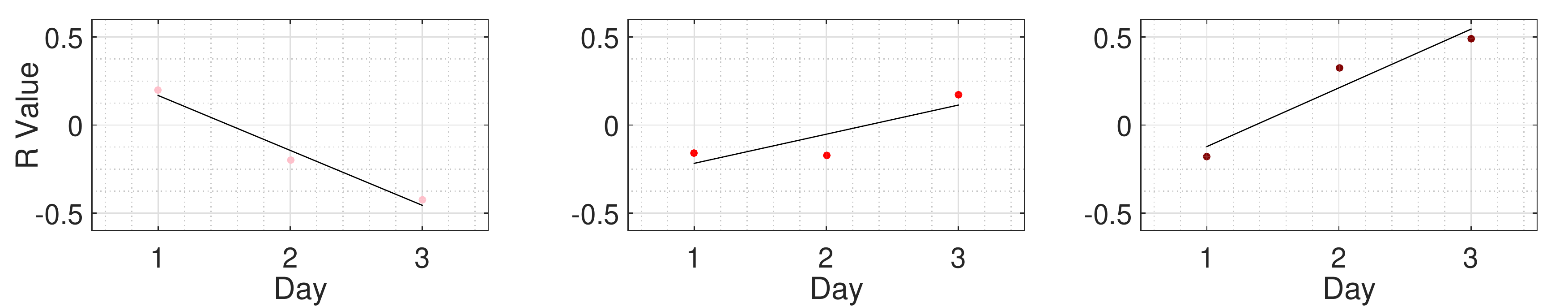}
	\caption{Scatter plots depicting the number of trials (level of task familiarity) and R values, defined here as correlations between learning rate parameters and the norm of the decomposed signal of interest (Pink points in the Left: $\bbx_\L$, Red points in the Middle: $\bbx_\M$, and Maroon points in the Right: $\bbx_\HH$). Top row: 6 week experiment with number of trials described in linear scale. Middle row: 6 week experiment withe number of trials evaluated in logarithm scale. We examine 6 week experiment by ordering the number of trials in both linear and logarithm scales to alleviate the fact that number of trials are densely distributed towards small values. Bottom row: 3 day experiment with scanning session ordered by the number of days in the experiment.}
	\label{fig_Freq_LR}
\end{figure*}

We evaluate the learning rate for all participants at each scanning session, and then compute the correlation between the norm $\|\bbx_\L\|_2$ of the decomposed signal corresponding to low spatial variation and the learning rates across subjects. The correlation (R value) between the norms $\|\bbx_\M\|_2$ as well as $\|\bbx_\HH\|_2$ and learning rates are also calculated. Fig. \ref{fig_Freq_LR} plots the Pearson correlation coefficients at all scanning sessions of the two experiments considered. The horizontal axis denotes the level of exposure of participants to the sequence -- which day in the 3 day experiment and how many number of trials participants have completed at the end of the scanning session in the 6 week experiment. Points are densely distributed for small number of trials in the 6 week experiment, so to mitigate this effect, we also plot the points by taking the logarithm of numbers of trials completed. We emphasize that due to normalization at each sampling point, the correlation values would all be $0$ if graph frequency decomposition segments brain signals into three equivalent pieces. There are scan sessions where the correlation is of particular interest, however the most noteworthy observation is the change of correlation values with the level of exposure of participants. For $\bbx_\L$ corresponding to smooth spatial variation, its correlation with learning is above zero ($\approx 0.25$) at the start of the training when participants perform the task for the first time. The correlation gradually decreases with heavier training intensity until below zero ($\approx -0.25$) at the end of the experiment when individuals are highly familiar with the sequence. For $\bbx_\HH$ corresponding to vibrant spatial variation, its correlation with learning is below zero ($\approx -0.2$) at the start of the training, and gradually increases throughout training until it is above zero ($\approx 0.25$) at the end of the experiment -- the exact opposite of $\bbx_\L$. For $\bbx_\M$, correlation between its norm $\|\bbx_\M\|_2$ with learning rate generally increases with the intensity of training, however this trend is not as obvious compared to other decomposition counterparts. The correlation between the number of trials and R value is summarized in Fig. \ref{fig_corr_R}. For robustness testing, we conduct similar analysis using the other two sets of parameters described in Section \ref{sec_network_illustration}. The average absolute difference for entries in Fig. \ref{fig_corr_R} between the original parameters and the other two parameter choices is $0.0377$. Again, similar observations are found in different experiments involving different learning tasks and different sets of participants.

%
\subsection{Discussion}\label{sec_learning_discussion}

This result further implies that the most association between learning or adaptability during the training process comes from the brain signals that either vary smoothly ($\bbx_\L$, regularity) or rapidly ($\bbx_\HH$, randomness) with respect to the brain network. Therefore, the graph frequency decomposition could be used to capture more informative brain signals by filtering out non-informative counterparts, most likely associated with middle graph frequencies. Besides, the positive association between $\|\bbx_\L\|_2$ and learning rates as well as the negative association between $\|\bbx_\HH\|_2$ and learning rates at the start of training indicates that it favors learning to have more {\it smooth}, {\it spread}, and {\it cooperative} brain signals when we face an unfamiliar task. As we gradually become familiar with the task, the smooth and cooperative signal distribution becomes less and less important, and there is a level of exposure when such signal distribution becomes destructive instead of constructive. We note that the task in the 3 day experiment is more difficult compared to that of the 6 week experiment, and therefore the time when the cooperative signal distribution starts to become detrimental (the point where the regression line intercepts the horizontal line of R value equaling $0$) is also comparable in the two experiments, describing a certain level of familiarity to the task. When we become highly familiar with the task, it is better and favors further learning to have {\it varied}, {\it spiking}, and {\it competitive} brain signals. 

In the dataset evaluated here, we utilize the average coherence between time series at pairs of brain cortical and subcortical regions during the training as the network. Hence, a concentration of brain activities towards low graph frequencies would imply that activities on brain regions that are generally cooperative are indeed similar. Simultaneously, the interpretation of concentration of brain activities towards high graph frequencies is that brain activities on brain regions that are generally cooperative are in fact dissimilar. In terms of learning, one possible explanation is that there are two different stages in learning: we start by grasping the big picture of the task to perform relatively well, and then we refine the details to perform better and to approach our limits.

Because the graph frequency analysis method presented in this paper applies to any setting where signals are defined on top of a network structure representing proximities between nodes, it would be interesting in future to use this method to investigate other types of signals and networks in neuroscience problems. As an example, in situations given fMRI measurements on structural networks, concentration of signals in low graph frequency components would imply functional activities do behave according to the structural networks.

Besides, it has been understood that learning is different when one is unfamiliar or familiar with a particular task -- it is easy to improve performance at first exposure due to the fact that one is far from their performance ceiling. It would therefore be interesting to utilize graph frequency decomposition to further analyze the difference between learning scenarios at different stages of familiarity, e.g. adaptability at first exposure and creativity when one fully understands the components of the specific tasks.

%
\section{Conclusion}\label{sec_conclusion}
We used graph spectrum methods to analyze functional brain networks and signals during simple motor learning tasks, and established connections between graph frequency with principal component analysis when the networks of interest denote functional connectivity. We discerned that brain activities corresponding to different graph frequencies exhibit different levels of adaptability during learning. Further, the strong correlation between graph spectral property of brain networks with the level of familiarity of tasks was observed, and the most contributing frequency signatures at different task familiarity was recognized.

%
\urlstyle{same}
\bibliographystyle{IEEEtran}
\bibliography{brain}

\end{document}